\documentclass{PoS}
\usepackage{amsfonts,wrapfig}
\usepackage{amsmath,dsfont}
\usepackage{amssymb}
\usepackage{multirow}
\usepackage{amsthm}
\usepackage{graphicx}

\newcommand{\p}[1]{\left(#1\right)}
\newcommand{\br}[1]{\left[#1\right]}

\newcommand{\davg}[1]{\left\langle #1\right\rangle}
\newcommand{\avg}[1]{\left\langle #1\right\rangle}

\newcommand{\be}[1]{\begin{equation} #1\end{equation}}

\title{$\rho$ meson decay on asymmetrical lattices}

\ShortTitle{$\rho$ meson decay LQCD}

\author{\speaker{Craig S. Pelissier},  Andrei Alexandru,  Frank X. Lee\\
        Department of Physics, The George Washington University, Washington, DC, 20052\\
        E-mail: \email{craigp@gwmail.gwu.edu}}

\abstract{
We present a lattice QCD calculation of the characteristics of the $\rho$ meson decay.  The study is carried out on spatially asymmetric boxes using nHYP-smeared clover fermions in the quenched approximation. The resonance mass and coupling constant are calculate using the P-wave scattering phaseshifts, of the isospin $I=1$ two-pion system. We use pion masses $m_{\pi}= 418$ MeV and $m_{\pi}=312$ MeV. In both cases, the $\rho$ decay is kinematically feasible. We work on lattice sizes $N_z\times24^2\times48$ with lattice spacing $a\approx0.1$ fm and $N_z=24,30,34,48$.
}

\FullConference{ The XXIX International Symposium on Lattice Field Theory - Lattice 2011\\
July 10-16, 2011\\
Squaw Valley, Lake Tahoe, California}

\begin{document}

\section{Introduction}

The study of multi-particle systems is an important step in understanding the strong dynamics of hadrons. In this report, we focus on the properties of the $\rho$ meson decay in the $I=1$ two-pion system. In the last year, several studies of the two-pion system on fully dynamical QCD gauge configurations have been done. These studies have taken steps to achieve a more reliable and precise calculation by using:  larger variational bases, better stochastic estimators, lower pion masses, larger boxes and the inclusion of strange dynamics \cite{Aoki:2011yj,Lang:2011mn,Feng:2010es,Frison:2010ws}. All these studies used Luscher's method \cite{Luscher:1991cf,Luscher:1990ux,Luscher:1990ck,Luscher:1986pf,Luscher:1985dn}, and its extensions to moving frames \cite{Rummukainen:1995vs}. In the following, we present results for the $\rho$ decay on 4-D asymmetrical tori in the center-of-mass (COM) frame.

To study the $\rho$ decay using current techniques, one must extract the two-pion spectrum in the energy regime where the $\rho$ decays. In order to scan this region, some mechanism to dial the relative momentum of the pions is required. In the moving frame formalism, the lattice calculation are carried out in a frame with non-zero total momentum. The results are then used to determine the two-pion spectrum in the COM. In this framework, the ability to tune the relative momentum of the pions is restricted because of the finite step size of the total momentum. In certain systems, this restriction may not allow the resonance region to be ideally mapped. A possible solution was proposed in which one or two of the spatial directions are elongated \cite{Li:2003jn}. In practice, a combination of the two methods would most likely lead to the most efficient calculation. In this work, we extract the $I=1$ two-pion P-wave scattering phaseshifts in the resonance region. The relative momentum of the two-pion system is adjusted by elongating $\textit{one}$ of the spatial directions. We use nHYP-smeared \cite{Hasenfratz:2007rf} clover fermions with $N_f=2$. The calculations are carried out in the quenched approximation using pion masses $m_{\pi}=312$ MeV and $m_{\pi}=418$ MeV.  We use lattice sizes  $N_z\times 24^2\times 48$ with lattice spacing $a\approx0.1$ fm and $N_z=24,30,34,48$.

The paper is organized as follows. In Section \ref{meth}, we discuss the details of the lattice calculation. The results are discussed in Section \ref{res}, and a summary is given in Section \ref{sum}.

\section{Methodology}\label{meth}

\begin{figure}[b]
\centering
\includegraphics[scale=0.45]{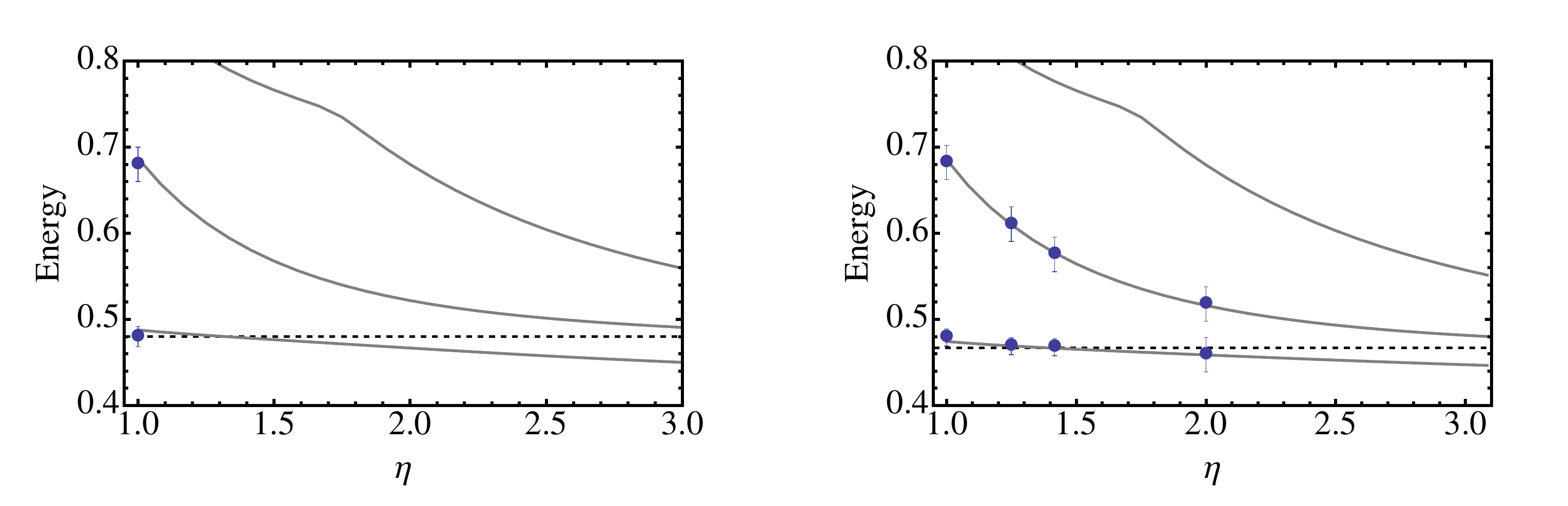}
\caption{In the left panel, the estimated energy spectrum is displayed for $m_{\pi}=418$ MeV. The blue dots are the calculated energies $E_0^{\eta=1}=0.48(1)$ and $E_1^{\eta=1}=0.68(2)$.The blue lines are the estimated behavior of the ground, first and second excited states. They are generated using $g_{\rho\pi\pi}=6.06$ and $am_{\rho}=0.48$. The value of $m_{\rho}$ is shown with a dashed line. From the figure, it can be seen that there is a relatively small discrepancy between $E_0^{\eta=1}$ and $m_{\rho}$. In the right panel, the final results are shown for two-pion spectrum with $m_{\pi}=418$ MeV.}
\label{energspec}
\end{figure}

In order to study the $\rho$ decay, we need to select appropriate lattice sizes. In this work, we consider lattice sizes with spatial volume $V_s=\eta 24 \times 24^2$ where we have introduce an elongation parameter $\eta$. The elongation of one spatial direction reduces the rotational symmetry to the the tetragonal group \cite{Hammermesh}. We work in the COM frame and fix the allowed angular momentum of the low-lying energies by projecting onto the one-dimensional $A_2^-$ representation. The projection restricts the relative momentum of the two-pion to the set
\begin{equation}
\left\{ \textbf{p}_{rel} \hspace{0.2cm}\left|\hspace{0.2cm}  \textbf{p}_{rel}=\left(0,0,\frac{2\pi}{\eta N_z}n_z, \right) \text{ for some } n_z\in\mathbb{Z} \text{ with } n_z\neq 0 \right\}\right.,
\end{equation}
and the invariant energy is
\begin{equation}
\sqrt{s} = 2\sqrt{m_{\pi}^2+p^2}
\end{equation}
The lowest angular momentum $l$ which couples to the $A_2^-$ channel is $l=1$. As a result, the zero momentum mode is excluded. To probe the resonance region, we need to select values of $\eta$ which will produce a ground state $E_0$ and, if possible, a first excited state $E_1$ which fall in the region where $\rho$ decays. In order to do this, we make predictions for the spectral behavior as a function of $\eta$ by enforcing agreement between the effective range formula
\begin{equation}
\cot{\delta}(s) = \frac{6\pi}{g^2_{\rho\pi\pi}} \frac{\sqrt{s}(m_{\rho}^2 - s)}{\p{\frac{s}{4}-m_{\pi}^2}^{3/2}},
\label{eff}
\end{equation}
and the relation between the scattering phaseshifts \cite{Li:2003jn}
\begin{equation}
\cot\delta(q)=\frac{1}{\pi^{3/2}\eta q}
\left\{  
\mathcal{Z}_{00}(1,q^2,\eta,1)+\frac{1}{q^2}\sqrt{\frac{2}{5}}\mathcal{Z}_{20}(1,q^2,\eta,1)
\right\}.
\label{luscher}
\end{equation}
where contributions from angular momentum $l>3$ have been ignored.
In Eq. \ref{eff}, $m_{\rho}$ is the resonance mass  and $g_{\rho\pi\pi}$ is the coupling constant defined through the effective Lagrangean
\begin{equation}
\mathcal{L}_{\text{eff}} = g_{\rho\pi\pi}\sum_{abc}\epsilon_{abc}(p_1-p_2)_{\mu}\rho^a_{\mu}(p)\pi^b(p_1)\pi^c(p_2).
\end{equation}
In Eq. [\ref{luscher}], $q=\frac{pL}{2\pi}$ with $p$ defined through the relation
\be{
p=\sqrt{\frac{s}{4}-m_{\pi}^2},
}
and $\mathcal{Z}_{lm}$ are generalized zeta functions whose definition can be found in \cite{Li:2003jn}. To generate an expectation for the spectrum, we need to fix $g_{\rho\pi\pi}$, $m_{\rho}$ and $m_{\pi}$. The pion mass is easily determined, and the coupling constant is taken to be the physical value $g_{\rho\pi\pi}$=6.06(1) . To estimate the resonance mass, we generate an ensemble with $\eta=1$. In this case,  the ground state is near the resonance mass, and we therefore make an initial guess $m_{\rho}=E_0^{\eta=1}$, refer to Fig. \ref{energspec}. Using the estimate for the low-lying spectrum, we make a conservative guess for the next value $\eta$. Subsequent values are selected by refining the values of $g_{\rho\pi\pi}$ and $m_{\rho}$ by fitting the available data with Eq. \ref{eff}.
To extract the low-lying energies, we use the variational method  \cite{Luscher:1990ck}. We construct a $2\times2$ matrix of correlation functions
\begin{equation}
C(t)_{ij}=\sum_n\langle n | \mathcal{O}^{\dag}_i(t)\mathcal{O}_j(0) | n \rangle\quad i,j=1,2.
\end{equation}
We use operators
\begin{eqnarray}
\mathcal{O}_1(t)\equiv \pi\pi(\textbf{p},t) &=&\frac{1}{\sqrt{2}}\left\{
\pi^-(\textbf{p},t)\pi^+(-\textbf{p},t)-\pi^+(\textbf{p},t)\pi^-(-\textbf{p},t)
\right\} \\
\mathcal{O}_2(t)\equiv \rho_3(\textbf{0},t) &=&\frac{1}{\sqrt{2}} 
\sum_{\textbf{x}}\left\{u(\textbf{x},t)\gamma_3\bar{u}(\textbf{x},t)-d(\textbf{x},t)\gamma_3\bar{d}(\textbf{x},t)\nonumber
\right\}
\end{eqnarray}
\begin{figure}[t]
\centering
\includegraphics[scale=0.45]{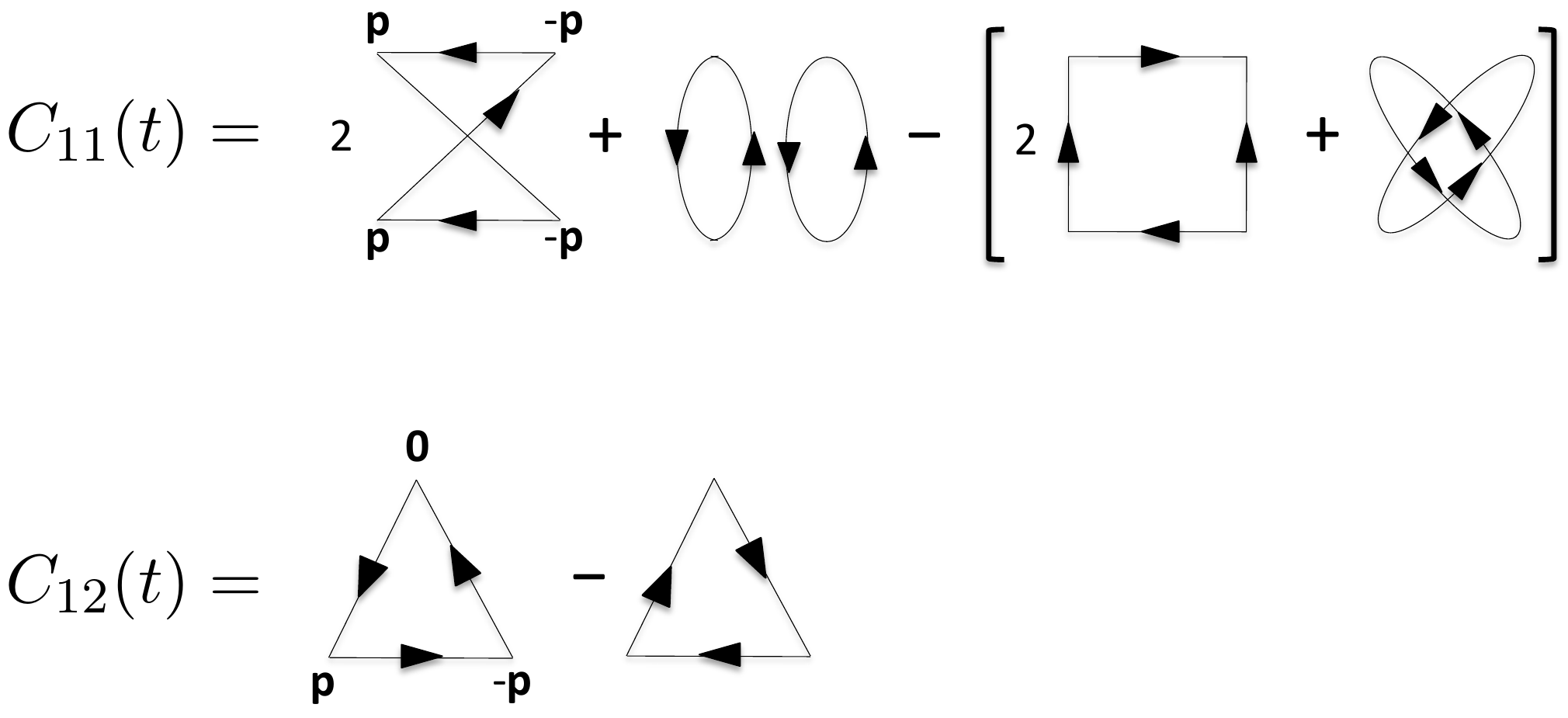}
\caption{Quark contractions for $C_{11}$ and $C_{12}$}.
\label{contractions}
\end{figure}%
where $\pi^{\pm}(\textbf{p},t)$ are interpolating operators for the $\pi^+$ and $\pi^-$ mesons with momentum $\textbf{p}=\frac{2\pi }{\eta L}\textbf{e}_3$.  To compute the correlation matrix, $C_{11}$ and $C_{12}$ need to be estimated stochastically. $C_{ 21}$ can be obtained though the relation $C_{21}=C^*_{12}$, and $C_{22}$ is evaluated by computing the quark propagator 
\begin{equation}
S_{AB}(\textbf{y},t | x_{source},t_{source}) 
\end{equation}
where $A$ and $B$ run over the 12 spin-color combinations. In Fig. \ref{contractions}, the quark contractions for $C_{11}$ and $C_{12}$ are shown. To compute them, we generate noise vectors with the property
\begin{equation}
\left<\xi^{\dag}\xi\right>=\mathds{1}_{\text{spatial}}\oplus\mathds{1}_{\text{spin}}\oplus\mathds{1}_{\text{color}}\label{noise}
\end{equation}
by placing an independently generated random $Z_4$ noise at each lattice point. In Eq. \ref{noise}, the angled brackets $\avg{\hdots}$ represent an average over the noise ensemble. Noise vectors are diluted by having support only on a single time slice and for a single spin-color combination. Introducing the lattice vectors
\begin{eqnarray}
u_{CD}(\textbf{x},t | \textbf{p} ,t_2,\xi_k) &=& \sum_{\textbf{y}}S_{CD}(\textbf{x},t | \textbf{y},t_2) e^{i\textbf{p}\cdot\textbf{y}}{\xi_k}(\textbf{y},t_2)\\
v_{AB}(\textbf{x},t | \textbf{p}_1,t_1| \textbf{p}_2, t_2 ,\xi_k ) &=& \sum_{\textbf{y},\textbf{z}}S_{AB}(\textbf{x},t | \textbf{y},t_1)\br{e^{i\textbf{p}_1\cdot\textbf{y}}\gamma_5S(\textbf{y},t_1 |\textbf{z},t_2)e^{i\textbf{p}_2\cdot\textbf{z}}\xi_k(\textbf{z},t_2)}
\end{eqnarray}
where $\xi^k$ represents the $k^{th}$ noise vector, the quark contractions for $C_{11}$ and $C_{12}$ can be constructed as shown in Table \ref{stoch}.
\begin{table}[t]
\centering
\label{stoch}
\begin{tabular}{c|c}
\multirow{4}{*}{\includegraphics[scale=0.8]{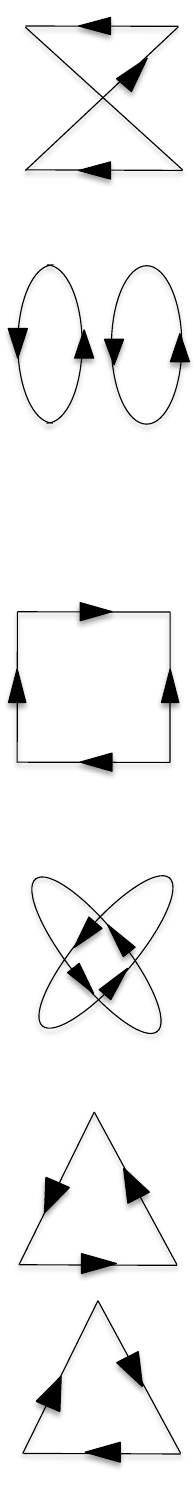}}&\\&$ \sum\limits_{\textbf{y}AB}e^{-i\textbf{p}\cdot\textbf{y}}\davg{v_{AB}(\textbf{y},t_f |- \textbf{p},t_i | \textbf{p} , t_i,\xi_k)v^{\dag}_{AB}(\textbf{y},t_f | -\textbf{p},t_f | \textbf{0},t_i,\xi_k) }$\\&\\
&$\br{\sum\limits_{\textbf{y}AB}e^{-i\textbf{p}\cdot\textbf{y}}\davg{u_{AB}(\textbf{y},t_f | \textbf{p} ,t_i,\eta_k)u^{\dag}_{AB}(\textbf{y},t_f | \textbf{0} ,t_i,\eta_k)}}$\\
&$\times\br{\sum\limits_{\textbf{y}AB}e^{i\textbf{p}\cdot\textbf{y}}\davg{u_{AB}(\textbf{y},t_f | -\textbf{p} ,t_i,\xi_k)u^{\dag}_{AB}(\textbf{y},t_f | \textbf{0} ,t_i,\xi_k)}}$\\
&\\&\\
&$\sum\limits_{\textbf{y}AB}e^{-i\textbf{p}\cdot\textbf{y}}\davg{v_{AB}(\textbf{y},t_f | \textbf{p},t_i | -\textbf{p} , t_i,\xi_k)v^{\dag}_{AB}(\textbf{y},t_f | -\textbf{p},t_f | \textbf{0},t_i,\xi_k) }$\\&\\
&$\br{\sum\limits_{\textbf{y}AB}e^{-i\textbf{p}\cdot\textbf{y}}\davg{u_{AB}(\textbf{y},t_f | -\textbf{p} ,t_i,\xi_k)u^{\dag}_{AB}(\textbf{y},t_f | \textbf{0} ,t_i,\xi_k)}}$\\
& $\times\br{\sum\limits_{\textbf{y}AB}e^{i\textbf{p}\cdot\textbf{y}}\davg{u_{AB}(\textbf{y},t_f | \textbf{p} ,t_i,\eta_k)u^{\dag}_{AB}(\textbf{y},t_f | \textbf{0} ,t_i,\eta_k)}}$\\&\\
&$\sum\limits_{\textbf{y}AB}\davg{v_{AB}(\textbf{y},t_f | \textbf{p},t_i | -\textbf{p} , t_i,\xi_k)u^{\dag}_{AB}(\textbf{y},t_f | \textbf{0} ,t_i,\xi_k)\gamma_5\gamma_3}$\\&\\
&$\sum\limits_{\textbf{y}AB}\davg{v_{AB}(\textbf{y},t_f | -\textbf{p},t_i | \textbf{p} , t_i,\eta_k)u^{\dag}_{AB}(\textbf{y},t_f | \textbf{0} ,t_i,\eta_k)\gamma_5\gamma_3}$\\
&
\end{tabular}
\caption{Stochastic estimators for the quark contractions of $C_{11}$ and $C_{12}$. Here $\xi$ and $\eta$ are independently generated noise ensembles and $\davg{\hdots}$ represents an average over the noise.}
\end{table}

\section{Results}\label{res}
\begin{figure}[b]
\centering
\includegraphics[scale=0.47]{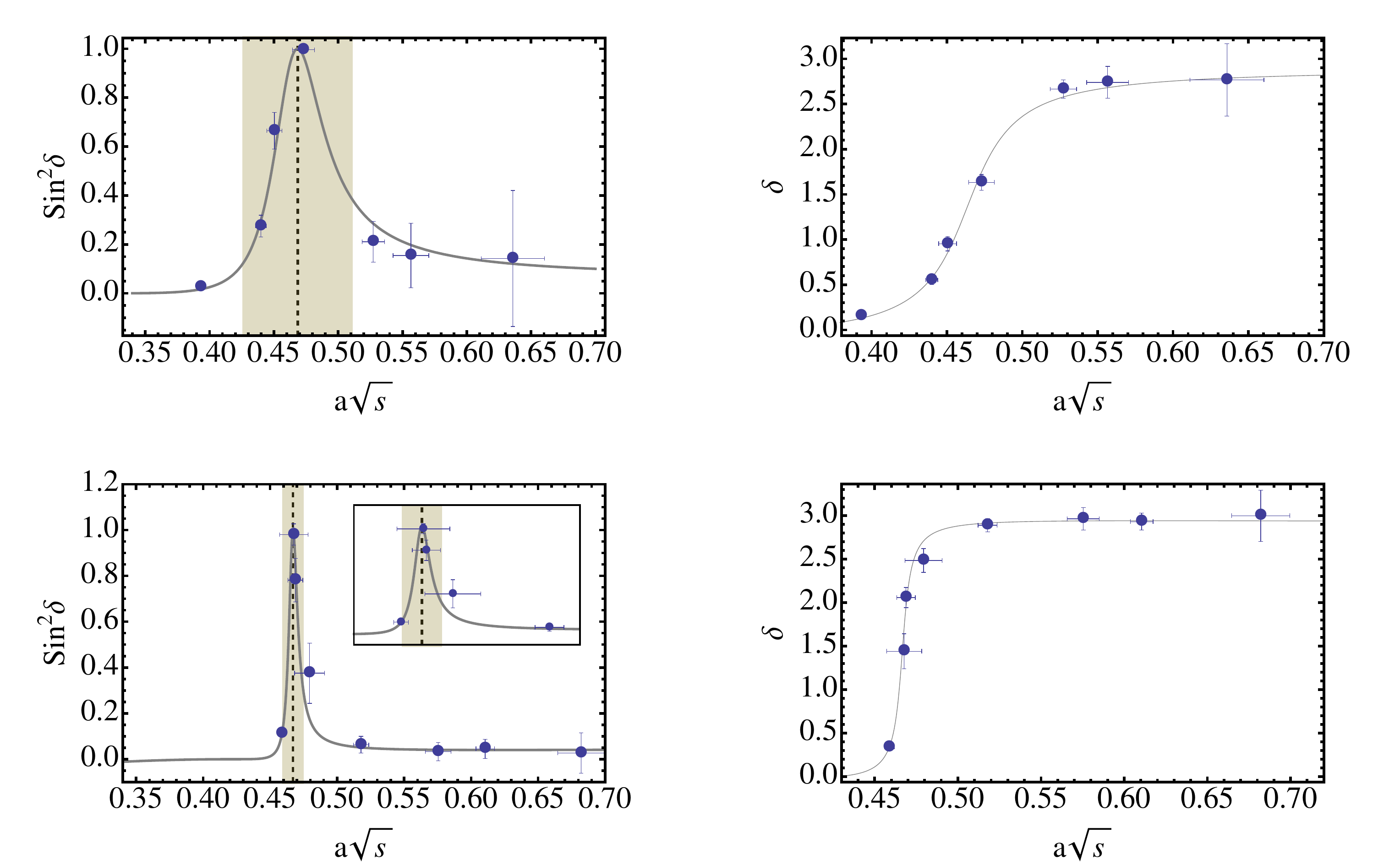}
\caption{Cross-section and phaseshift plots for $m_{\pi}=312$ MeV, upper panels, and $m_{\pi}=418$ MeV, lower panels. The dashed line marks the calculated value of $m_{\rho}$. In yellow, we shade the region $m_{\rho} - \Gamma \leq a\sqrt{s} \leq m_{\rho}+\Gamma$.} 
\label{plots}
\end{figure}
\begin{figure}
\centering
\includegraphics[scale=0.45]{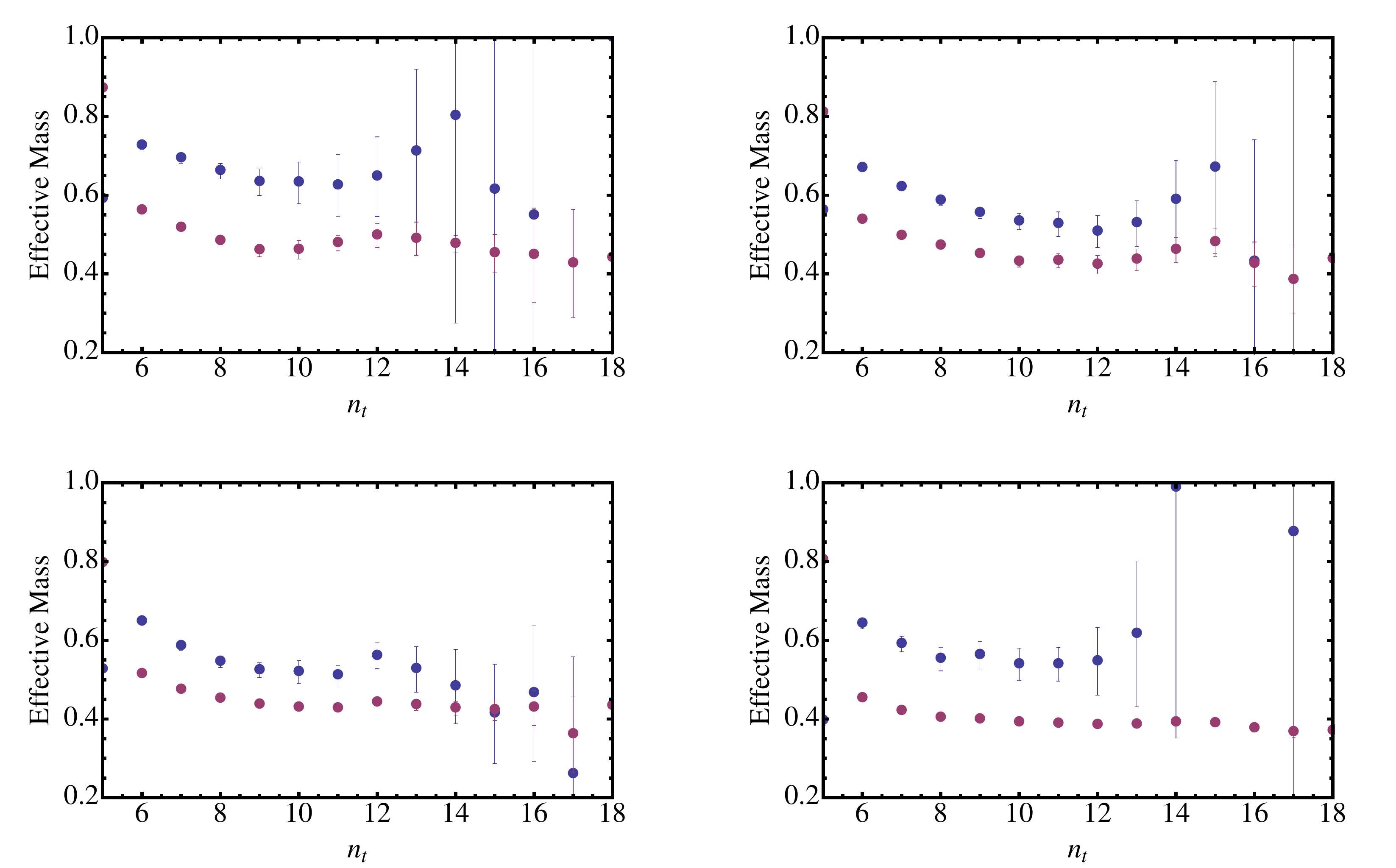}
\caption{Effective mass plots for $m_{\pi}=312$ MeV. The elongation parameter $\eta$ is $1.0,1.25,1.42$ and $2.0$ for the upper left, upper right, lower left and lower right panels, respectively.}
\label{eff310}
\end{figure}
\begin{figure}
\centering
\includegraphics[scale=0.45]{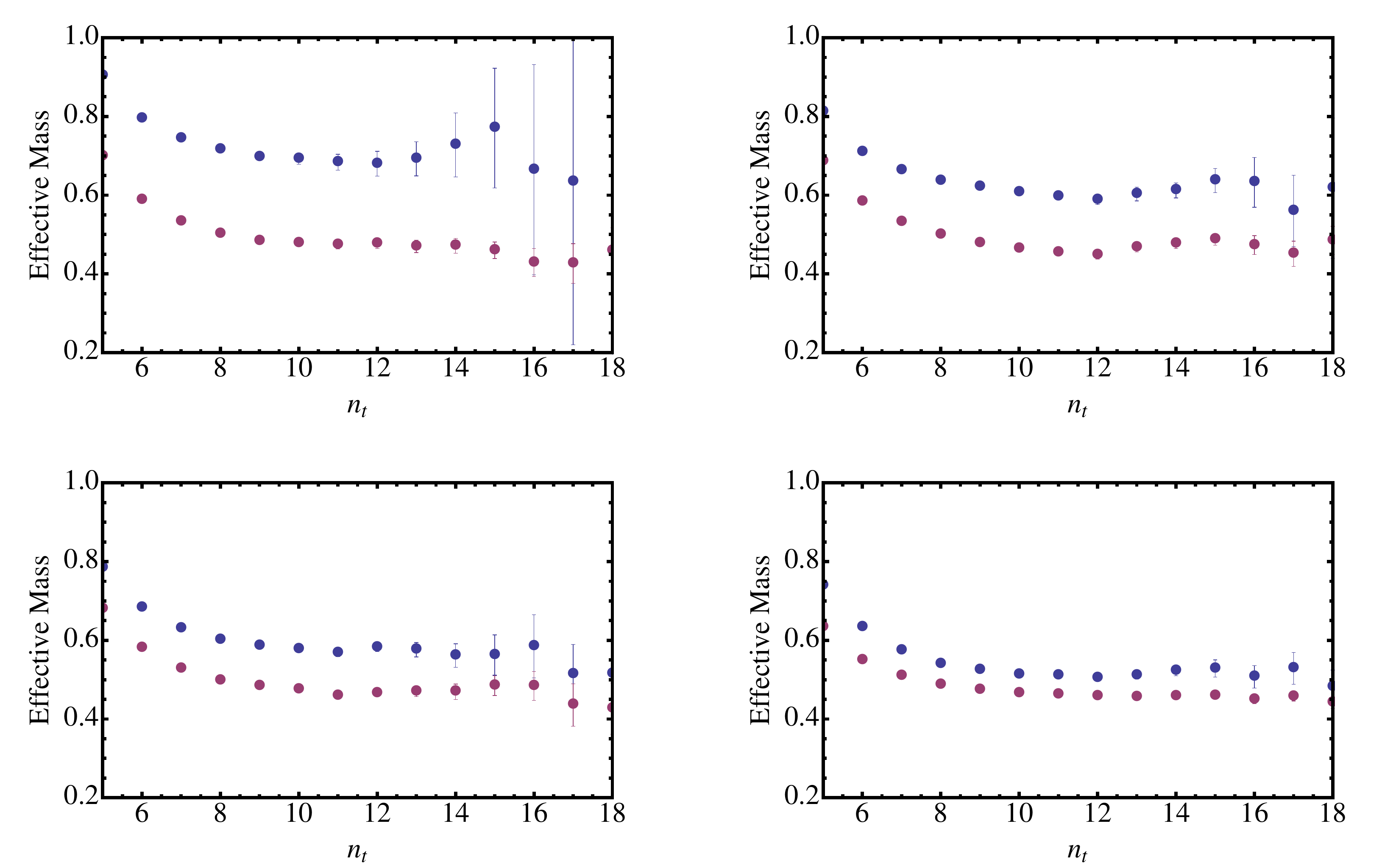}
\caption{Effective mass plots for $m_{\pi}=418$ MeV. The elongation parameter $\eta$ is $1.0,1.25,1.42$ and $2.0$ for the upper left, upper right, lower left and lower right panels, respectively.}
\label{eff418}
\end{figure}

In Figs. \ref{eff310} and \ref{eff418}, we show the effective mass plots for eigenvalues $\lambda_1(t,t_R)$ and  $\lambda_2(t,t_R)$ of the matrix $M(t,t_R)=C^{-1/2}(t_R)C(t)C^{-1/2}(t_R)$ with reference time $t_R=6$. The two-pion spectrum is calculated by fitting $\lambda_1(t,t_R)$ and  $\lambda_2(t,t_R)$ to the curve
\be{
\lambda(t)=c e^{\sqrt{s} N_t/2}\cosh\br{\sqrt{s}(N_t/2-t)}
}
in the temporal region where the largest energy is dominate. We compute the cross-section and phaseshifts from the spectral values through Eq. \ref{luscher}. The resonance mass $m_{\rho}$ and coupling constant $g_{\rho\pi\pi}$ are determined using non-linear regression with weights $w_i=\sigma_i^{-1}$ where $\sigma_i$ is the variance of the $i^{th}$ spectral value. The width of the resonance is computed using
\be{
\Gamma =\frac{g_{\rho\pi\pi^2}}{6\pi}\frac{\p{m_{\rho}^2/4-m_{\pi}^2}^{3/2}}{m_{\rho}^2}.\label{width}
}
We estimate the physical value $\Gamma_{ph}$ by evaluating Eq. \ref{width} with the extracted values of $g_{\rho\pi\pi}$ and the physical values of $m_{\rho}$ and $m_{\pi}$. The results are listed in Table \ref{final-results} and displayed in Fig. \ref{plots}. For the two pion masses, we found coupling constants which agree within error and are consistent with the physical value $g_{\rho\pi\pi}=6.06(1)$. The resonance mass showed no appreciable quark mass dependence. The only substantial change was a widening of the resonance at the lower pion mass, refer to Fig. \ref{plots}.
\begin{table}[t]
\centering
\label{final-results}
\begin{tabular}{c|c|c|c|c}
\hline\hline
$m_{\pi}$ [MeV]  & $m_{\rho}$ [MeV] & $g_{\rho\pi\pi}$ & $\Gamma_{ph}$ [MeV] & $\Gamma$ [MeV]\\
\hline\hline
312(4) & 924(4)  & 5.85(35) & 142(16)&85(8)\\
418(2) &  922(1) & 6.25(27) & 161(14)&16(2)\\
\hline\hline
\end{tabular}
\caption{Fit results for the resonance mass $m_{\rho}$, coupling constant $g_{\rho\pi\pi}$ and decay width $\Gamma$. The physcal decay width $\Gamma_{ph}$ is computed using the calculated value for $g_{\rho\pi\pi}$ and the physical values of $m_{\rho}$ and $m_{\pi}$.}
\end{table}%
Despite working in the quenched approximation, our results are similar to studies done on fully dynamical QCD gauge configurations. A reasonable comparison can be made by looking at the results of \cite{Aoki:2011yj}, since both studies used $O(a)$ improved wilson fermions with similar lattice spacings and comparable values of $m_{\pi}$ and $m_{\rho}$. In both cases, the coupling constant $g_{\rho\pi\pi}$ was  consistent with its physical value and did not show significant quark masses dependence. The resonance mass $m_{\rho}$ was also found to have a non-substantial quark mass dependence. This suggest that the quenched approximation is able to capture qualitatively, the physical dynamics of the $\rho$ decay. However, a proper comparison would require studies which only differ by quenching effects.

\section{Summary}\label{sum}

In this paper, we discussed the calculation of the $\rho$ decay on spatially asymmetric 4-D tori.  By varying the degree of asymmetry, we were able to calculate several well separated phaseshifts in the region where $\rho$ decays. The study was carried out in the quenched approximation and showed comparable results with similar studies done on fully dynamical QCD configurations. We are currently performing the calculation on fully dynamical QCD configurations. This will give a more accurate picture of the quenching effects and allow for a comparison with the moving frame formalism used in other studies.

\section{Acknowledgements}
This work is supported in part by the U.S. Department of Energy grant
DE-FG02-95ER-40907 and GW IMPACT collaboration.

\bibliographystyle{JHEP}
\bibliography{citations}

\end{document}